\documentclass[preprint,showpacs,preprintnumbers,amsmath,amssymb]{revtex4}
\usepackage{graphicx}
\usepackage{dcolumn}
\usepackage{bm}
\usepackage{epsfig}
\begin{document}
\title{Ferromagnetism of electron gas}

\author{H.Gholizade}\email{gholizade@ut.ac.ir} \affiliation{Engineering Science Department,
University of Tehran, Enghelab Sq., Tehran, Iran.}
\author{D.Momeni}\affiliation{ Department of physics,Faculty of basic
sciences,Tarbiat Moalem university,Karaj,IRAN}

\begin{abstract}
In current work, we investigate the density and temperature
dependence of polarization parameter; using the relativistic
formalism for the electron-electron interaction within the fermi
liquid model. we calculate the spin dependent scattering matric
elements in relativistic region, and then obtain the
non-relativistic behavior to study the magnetic properties of an
electron gas. By varying the polarization parameter, we minimized
the free energy and then obtain the polarization of the system as a
function of density and temperature. At zero temperature the exact
results for polarization and magnetic susceptibility have obtained.
It has been shown that for a given temperature (density) there is
critical density (temperature) that the ferromagnetic phase can
appears in electron gas. Our results show that at nonzero
temperatures and in very low and very hight densities the
ferromagnetism phase can not be exist.
\end{abstract}
 \maketitle
\section{\label{sec:level1}Introduction}
The properties of electron gas are investigated in several works
\cite{landau,kwon}. One of the important cases, is the
ferromagnetism of electron gas \cite{landau}. The spontaneous
magnetization may appear in different densities for different
temperatures and the polarization of the system is a function of
density and temperature. By assuming the spin-spin interaction
inside the system we can study the magnetic property of the system.
Several models study the possibility of existence of ferromagnetism
phase inside the solids by assuming the lattice system
\cite{landau}. For gas systems by using the statistical methods for
an imperfect fermi gas we can show that the ferromagnetism phase can
be exist \cite{landau}. In current work we examine the possibility
of existence of ferromagnetism phase inside the fermi gas by one
photon exchange interaction between two electron and Fermi liquid
model to calculate the energy density of the system. Because the
spin dependence of the Landau Fermi liquid interaction function is
due to the relativistic effects (spin-spin and spin-orbit
interaction) and the exchange interaction \cite{landau}, we
calculate the direct and exchange diagrams contributions (with
defined spin directions) to thermodynamic quantities, in
relativistic formalism. After this we obtain the non-relativistic
limit,because in the ultra-relativistic region the exchange
interaction becomes repulsive and can not cause to ferromagnetic
phase inside system. By varying the free energy with respected to
the p(polarization parameter) at various density and temperature we
can find the minimum of the free energy for given density and
temperature. At end, we obtained the magnetic susceptibility of
system at zero temperature, and our results shoe that the $\chi$ has
a divergence. This means that the phase transition is second order
inside the system.

\section{Exchange and direct contributions in energy density}
To obtain the exchange and direct diagrams contributions we write
the Lorentz invariant matric elements for two electron scattering in
order $g^2$. It can be shown that the direct diagram portion in
energy density is \cite{huang}
\begin{equation}
E_d\sim -\rho V^{\frac{2}{3}}, \label{1}
\end{equation}
the $\frac{E_d}{N}\sim V^{\frac{-1}{3}}$ becomes zero in the
thermodynamic limit. So we can ignore the direct interactions and
only calculate the exchange interactions contribution in the
thermodynamic limit. We define the up and down spin states in the
rest frame of each electron, with the vector $\vec{\xi}=(0,0,\pm1)$.
Where the up and down spin states corresponded to the $\pm$ signs.
Electrons with momentum $p$ and spin $s$ are described by spinors
$u(p,s)$, where the $s^\mu$ is a Lorentz vector. In the rest frame
of particles the $s^\mu$ reduced to the unit spatial vector
$\vec{\xi}=(0,0,\pm1)$. The components of $s^\mu$ in the frame which
the particle moves with momentum $\vec{p}$ are obtained by Lorentz
transformation \cite{tatsumi}:
\begin{equation}
s^\mu=(\frac{\vec{p}.\vec{\xi}}{m},\vec{\xi}+\frac{\vec{p}.\vec{\xi}}{m(E+m)}\vec{p}).
\label{2}
\end{equation}
Where $E=\sqrt{m^2+p^2}$. It can be checked that the Eq(\ref{2})
satisfied the normalization and orthogonality conditions, which in
the rest frame are evidence. State with definite polarization
obtained by applying the projection operator
$\Sigma(s)=\frac{1}{2}(1+\gamma_5\slash \!\!\!s)$. The polarization
density matrix can be written as\cite{tatsumi}:
\begin{equation}
\rho(p,s)=\frac{1}{2m}(\slash \!\!\!\! p+m)\Sigma(s). \label{3}
\end{equation}
For exchange interaction we can write the Lorentz invariant matrix
element as follow \cite{tatsumi}:
\begin{eqnarray}
{\cal M}^s_{{\bf k}\xi,{\bf q}\xi'}= g^2{\rm tr}(\gamma_\mu
\rho(k,\xi) \gamma^\mu \rho(q,\xi'))\frac{1}{(k-q)^2}. \label{4}
\end{eqnarray}
It must mentioned that if we use above relation in Landau theory,
then the stability of the fermi liquid do not satisfied. If we add a
term like $\delta ^2$ in the dominator of the Eq.(\ref{4}), then the
stability can satisfied. The $\delta ^2$ can interpreted as higher
order corrections to the gauge boson propagator. For simplicity we
ignore the $\delta^2$ and set it equal to zero. After performing
traces the result becomes \cite{tatsumi}:
\begin{eqnarray}
{\cal M}^s_{{\bf k}\xi,{\bf
q}\xi'}=\frac{1}{2}\frac{g^2}{m^2}[2m_q^2-k.q-(\textbf{k}
.\xi)(\textbf{q}.\xi')\nonumber\\
&&\hspace{-50mm}m^2\xi.\xi'+\frac{1}{(\varepsilon_k+m)(\varepsilon_q+m)}\nonumber\\
&&\hspace{-46mm}\times\{m(\varepsilon_k+m)(\xi.\textbf{q})(\xi'.\textbf{q})+\nonumber\\
&&\hspace{-40.5mm}m(\varepsilon_q+m)(\xi.\textbf{k})(\xi'.\textbf{k})+\nonumber\\
&&\hspace{-40mm}(\textbf{k.q})(\xi.\textbf{k})(\xi'.\textbf{q})\}]\frac{1}{(k-q)^2}
\label{5}
\end{eqnarray}
The Landau Fermi liquid interaction function related to Lorentz
invariant matrix element via:
\begin{equation}
f_{k\xi,q\acute{\xi}}=\frac{m^2}{E_kE_q}{\cal M}^s_{{\bf k}\xi,{\bf
q}\xi'} \label{6}
\end{equation}
If $\xi=\xi'$ (parallel spins) we have the spin non-flip interaction
and if $\xi=-\xi'$ (anti parallel spins) we have flip interaction.
So the exchange energy density for flip and non-flip interactions
can be written:
\begin{eqnarray}
\varepsilon_{ex}^{flip}=\int\int\frac{d^3k}{(2\pi)^3}\frac{d^3q}{(2\pi)^3}
n(k^+)n(q^-)f_{{\bf k},{\bf q}}^{flip} \label{7}
\end{eqnarray}
\begin{eqnarray}
\varepsilon_{ex}^{non-flip}=\frac{1}{2}\sum_{i=\pm}\int\int
\frac{d^3k}{(2\pi)^3}\frac{d^3q}
{(2\pi)^3}\times\nonumber\\
&&\hspace{-30mm}n(k^i)n(q^i)f_{{\bf k},{\bf q}}^{non-flip} \label{8}
\end{eqnarray}
In above equations $\pm$ correspond to
\begin{eqnarray}
n_+=n_q(1+p)/2\nonumber\\
n_-=n_q(1-p)/2 \label{9}
\end{eqnarray}
Where the $n_{\pm}$ and $p$ are the density of spin up and spin down
quarks and polarization parameter, respectively. $n(k^i)$ are the
Fermi distribution functions. We can calculate integrals over angles
using following relations \cite{mathews}:
\begin{eqnarray}
\int\frac{d\Omega}{1+\vec{k}.\hat{r}}=\frac{2\pi}{k}\ln(\frac{1+k}{1-k})\nonumber\\
&&\hspace{-50mm}
\int d\Omega (\hat{r}.\vec{a})(\hat{r}.\vec{b})=\frac{4\pi}{3}\vec{a}.\vec{b}\nonumber\\
&&\hspace{-50mm} \int d\Omega
\frac{\vec{a}.\hat{r}}{1+\vec{k}.\hat{r}}=\frac{4\pi}{k^2}
\vec{a}.\vec{k} [1-\frac{1}{2k}\ln(\frac{1+k}{1-k})]\nonumber\\
&&\hspace{-50mm} \int
d\Omega\frac{(\vec{a}.\hat{r})(\vec{b}.\hat{r})}{1+\vec{a}.\hat{r}}
=\frac{2\pi}{a^3}\ln(\frac{1+k}{1-k})\vec{a}.\vec{b} \label{10}
\end{eqnarray}
After integrating over angles we perform the integration on the
momentums by numerical methods. The numerical integration arguments
are as follow:
\begin{eqnarray}
\varepsilon_{ex}=\frac{1}{(2\pi)^6}\int k^2q^2
dkdq\frac{A(k,q)}{B(k,q)}(\sum_{i=1}^{6}T_i) \label{11}
\end{eqnarray}
with:
\begin{eqnarray}
&&\hspace{-10mm}A(k,q)=\frac{m^2}{\varepsilon_k\varepsilon_q}\frac{2g^2}{18m^2kq}\nonumber\\
&&\hspace{-10mm}
B(k,q)=\frac{m^2-\varepsilon_k\varepsilon_q}{kq}\nonumber\\
&&\hspace{-10mm}
B_1(k,q)=2m^2+m^2\vec{\zeta}.\vec{\zeta'}-\varepsilon_k\varepsilon_q\nonumber\\
&&\hspace{-10mm}
Z_1(k,q)=\frac{m(\varepsilon_k+m)}{(\varepsilon_k+m)(\varepsilon_q+m)}\nonumber\\
&&\hspace{-10mm}
Z_2(k,q)=\frac{m(\varepsilon_q+m)}{(\varepsilon_k+m)(\varepsilon_q+m)}\nonumber\\
&&\hspace{-10mm}
Z_3(k,q)=\frac{1}{(\varepsilon_k+m)(\varepsilon_q+m)}\nonumber\\
&&\hspace{-10mm}
T_1=8\pi^2B_1(k,q)\ln(\frac{B(k,q)+1}{B(k,q)-1})\nonumber\\
&&\hspace{-10mm}
T_2=16\pi^2kq(1-\frac{B(k,q)}{2}\ln(\frac{B(k,q)+1}{B(k,q)-1}))\nonumber\\
&&\hspace{-10mm} T_3=\frac{-16\pi^2}{3}k
q(1-\frac{B(k,q)}{2}\ln(\frac{B(k,q)+1}{B(k,q)-1}))
\vec{\zeta}.\vec{\zeta'}\nonumber\\
&&\hspace{-10mm}
T_4=\frac{8\pi^2}{3}Z_1(k,q)q^2\ln(\frac{B(k,q)+1}{B(k,q)-1})\vec{\zeta}.
\vec{\zeta'}\nonumber\\
&&\hspace{-10mm}
T_5=\frac{8\pi^2}{3}Z_2(k,q)k^2\ln(\frac{B(k,q)+1}{B(k,q)-1})\vec{\zeta}.
\vec{\zeta'}\nonumber\\
&&\hspace{-10mm}
T_6=Z_3(k,q)\frac{8\pi^2}{3}B(k,q)^2k^2q^2\ln(\frac{B(k,q)+1}{B(k,q)-1})
\vec{\zeta}.\vec{\zeta'} \label{12}
\end{eqnarray}
Integration results for $p=0$ at zero temperature are the same as
unpolarized exchange energy results that obtained before
\cite{zapolsky,gordonbaym}.
\begin{eqnarray}
\varepsilon_{ex}^{unpol} = -\frac{\alpha}{4\pi^3}\{k_F^4&
&\nonumber\\&&\hspace{-20mm} - \frac{3}{2}[
{E_Fk_F-m^2_q\ln(\frac{E_F+k_F}{m_q})}]^2\}, \label{13}
\end{eqnarray}
where   $E_F=\sqrt{m_q^2+k_F^2}$ is the Fermi energy
\cite{zapolsky,gordonbaym,akhiezer} and the $\alpha=g^{2}/4\pi$.
According to fermi liquid theory the interaction function on the
Fermi surface has the form\cite{landau}:
\begin{equation}
(\frac{k_f^2}{\pi^2
v_f})f_{k\xi,q\acute{\xi}}=F(\theta)+\vec{s}.\vec{s'}G(\theta).
\label{14}
\end{equation}
In Eq.(\ref{14}) $\theta$ is the angle between two electron momentum
on the Fermi surface. Comparing Eq.(\ref{14}) with Eq.(\ref{6}) and
Eq.(\ref{5}) one can see that on the Fermi surface only the scalar
product of two electron spin operators appears \cite{landau}. In
non-relativistic region we can use the $k_f\ll m$ approximation and
then we have:
\begin{equation}
{\cal M}_{non-rel}=\frac{-g^2}{2}
\frac{1+\vec{\xi}.\vec{\xi'}}{|\vec{k}-\vec{q}|^2} \label{15}
\end{equation}
One can see that if we have $\xi=-\xi'$ then the Lorentz invariant
matrix elements vanishes, this means that in non-relativistic region
the spin flip contribution in interaction energy density vanishes.
The spin non-flip exchange and kinetic energy density in
non-relativistic case at zero temperature is:
\begin{eqnarray}
\varepsilon_{ex}^{non-rel}(T=0)=\frac{-\alpha}{8\pi^3}k_f^4
((1+p)^{\frac{4}{3}}+(1-p)^{\frac{4}{3}})\nonumber\\
\varepsilon_{kin}^{non-rel}(T=0)=\frac{k_f^5}{20m\pi^2}
((1+p)^{\frac{5}{3}}+(1-p)^{\frac{5}{3}}) \label{16}
\end{eqnarray}
\section{Equation of state at low temperature}
In the low temperature the kinetic and exchange energy densities,
and entropy density of the system become:
\begin{eqnarray}
\varepsilon_{kin}^{non-rel}(T)=\varepsilon_{kin}^{0}
[1+\frac{5\pi^2}{12}(\frac{2m^{*}T}{k_f^2})^2]\nonumber\\
\varepsilon_{ex}^{non-rel}(T)=\varepsilon_{ex}^{0}
[1-\frac{\pi^2}{6}(\frac{2m^{*}T}{k_f^2})^2]\nonumber\\
S=s^++s^-=\frac{\pi^2nm^*T}{2k_f^2}[(1+p)^{\frac{1}{3}}+(1-p)^{\frac{1}{3}}]
\label{17}
\end{eqnarray}
where the $m^*$, $\varepsilon_{kin}^{0}$ and $\varepsilon_{ex}^{0}$
are the effective mass of electrons, non-relativistic kinetic and
exchange energies at zero temperature, respectively. Using the
results of equation (\ref{17}) one can obtain the following result
for free energy density:
\begin{eqnarray}
F=\varepsilon_{ex}+\varepsilon_{kin}-TS\nonumber\\
=\varepsilon_{ex}^{0}
[1-\frac{\pi^2}{6}(\frac{2m^{*}T}{k_f^2})^2]\nonumber\\
+\varepsilon_{kin}^{0}
[1+\frac{5\pi^2}{12}(\frac{2m^{*}T}{k_f^2})^2]\nonumber\\
-\frac{\pi^2nm^*T}{2k_f^2}[(1+p)^{\frac{1}{3}}+(1-p)^{\frac{1}{3}}]
\label{18}
\end{eqnarray}
According to the thermodynamics fundamental relations, the free
energy must be minimum at the equilibrium state, so we must have:
\begin{equation}
dF|_{T,n}=0 \label{19}
\end{equation}
At fixed density and temperature we can write:
\begin{equation}
\frac{\partial F}{\partial m^*}dm^*+\frac{\partial F}{\partial
p}dp=0 \label{20}
\end{equation}
Because the effective mass and polarization parameter are the
independent variables for free energy, then we must have:
\begin{eqnarray}
\frac{\partial F}{\partial m^*}=0\nonumber\\
\frac{\partial F}{\partial p}=0 \label{21}
\end{eqnarray}
Solving the above equations simultaneously, we can obtain the
density and temperature dependence of $m^*$ and $p$. The derivatives
of free energy with respected to $p$ and $m^*$ can be written as
follow:
\begin{eqnarray}
\frac{\partial F}{\partial m^*}
=(\frac{T}{k_f^2})^2m^*[\varepsilon_{kin}^{0}(\frac{10\pi^2}{3})
-\varepsilon_{ex}^{0}\frac{4\pi^2}{3}]\nonumber\\
-(\frac{T}{k_f})^2\frac{\pi^2n}{2}[(1+p)^{\frac{1}{3}}+(1-p)^{\frac{1}{3}}]
\label{22}
\end{eqnarray}
\begin{eqnarray}
\frac{\partial F}{\partial
p}=[1+\frac{5\pi^2}{12}(\frac{2m^*T}{k_f}^2)]\frac{\partial
\varepsilon_{kin}^{0}}{\partial
p}\nonumber\\
+[1-\frac{\pi^2}{6}(\frac{2m^*T}{k_f})^2]\frac{\partial
\varepsilon_{ex}^{0}}{\partial p}\nonumber\\
-\frac{\pi^2nm^*}{6}(\frac{T}{k_f})^2[(1+p)^{\frac{1}{3}}-(1-p)^{\frac{1}{3}}]
\label{23}
\end{eqnarray}
The derivatives of kinetic and exchange energies with respected to
$p$ are given bellow:
\begin{eqnarray}
\frac{\partial}{\partial p} \varepsilon_{ex}^{0}=-\frac{4}{3}
\frac{\alpha}{8\pi^3}k_f^4[(1+p)^{\frac{1}{3}}-(1-p)^{\frac{1}{3}}]\\
\frac{\partial}{\partial p} \varepsilon_{kin}^{0}=\frac{5}{3}
\frac{k_f^5}{20\pi^2m}[(1+p)^{\frac{2}{3}}-(1-p)^{\frac{2}{3}}]
\label{24}
\end{eqnarray}
By solving the equation (\ref{22}), one can find $m^*$ as a function
of $p$.
\begin{equation}
m^*=\frac{4k_f^2
n}{3}\frac{(1+p)^{\frac{1}{3}}+(1-p)^{\frac{1}{3}}}{5\varepsilon_{kin}^{0}
-2\varepsilon_{ex}^{0}} \label{25}
\end{equation}
Inserting this function in equation (\ref{23}), we can obtain the
density and temperature dependence of polarization parameter. At
zero temperature the equation (\ref{23})becomes simpler and one can
obtain the bellow result:
\begin{equation}
\frac{\partial}{\partial p}
\varepsilon_{ex}^{0}+\frac{\partial}{\partial p}
\varepsilon_{kin}^{0}=0
\label{26}
\end{equation}
The resultant equation from Eq.(\ref{26}) is:
\begin{eqnarray}
\frac{k_f}{2m}[(1+p)^{\frac{2}{3}}-(1-p)^{\frac{2}{3}}]=\frac{\alpha}{\pi}
[(1+p)^{\frac{1}{3}}-(1-p)^{\frac{1}{3}}] \label{27}
\end{eqnarray}
This result is very look like to the well known results of the
spontaneous magnetization of an imperfect Fermi gas \cite{huang}.
But in our calculus the interaction part of the hamiltonian is not
independent of the spins alignment, and this dependence change the
right side of the equation (\ref{27}).
\section{Results and Discussion}
\begin{figure}[htb]
\epsfxsize=10.5cm \epsfbox{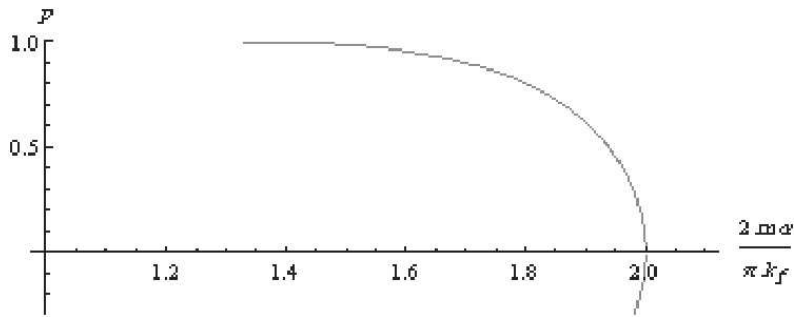} \caption{The polarization
parameter of system at zero temperature. The horizontal axis shows
the $\frac{2\alpha m}{\pi k_f}$. For $\alpha m = \pi k_f $ the
polarization of the system vanishes. The value of critical density
for phase transition is proportional to the coupling constant. If we
ignore the interaction inside system the ferromagnetism can not
appeared and if we use the effective and great coupling constant
then the critical density increases. } \label{f1}
\end{figure}
\begin{figure}[htb]
\epsfxsize=10.5cm \epsfbox{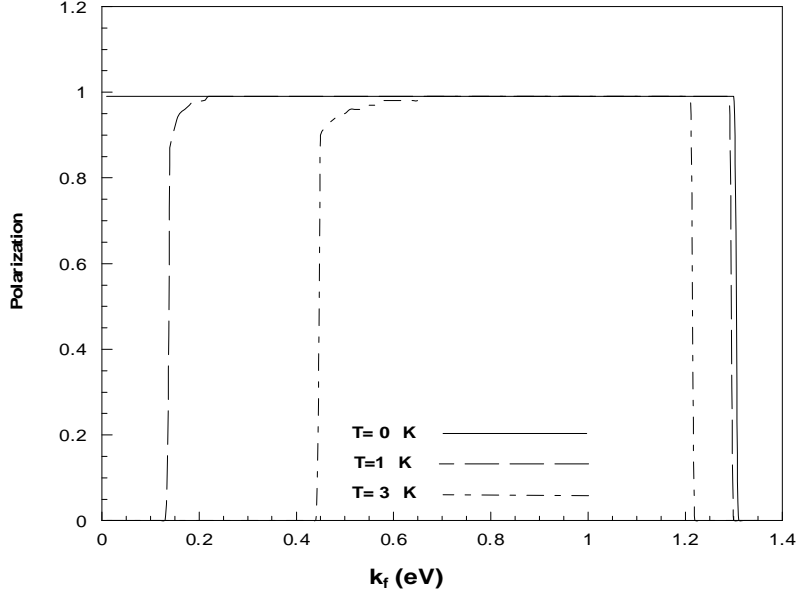} \caption{The density dependence of
polarization parameter at various temperatures. By increasing the
temperature the ferromagnetism bound of the system become narrower.
So we expect that for specific temperature ($T_\alpha$) the
ferromagnetic phase completely disappeared and system can not be in
ferromagnetic phase at any densities. The $T_\alpha$ strongly
depended on interactions inside the system or in other words
depended strongly on coupling constant. } \label{f2}
\end{figure}
\begin{figure}[htb]
\epsfxsize=10.5cm \epsfbox{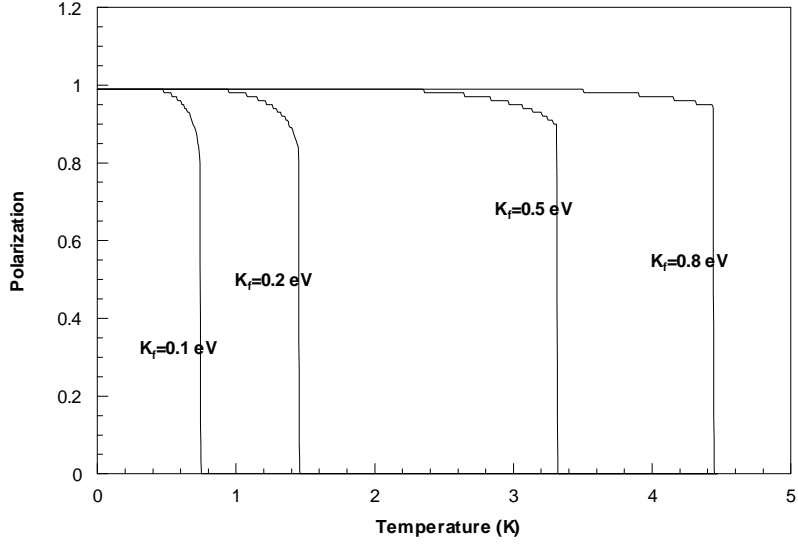} \caption{The polarization
parameter as a function of temperature at different densities. By
increasing temperature, the polarization of system reduced.}
\label{f3}
\end{figure}
\begin{figure}[htb]
\epsfxsize=10.5cm \epsfbox{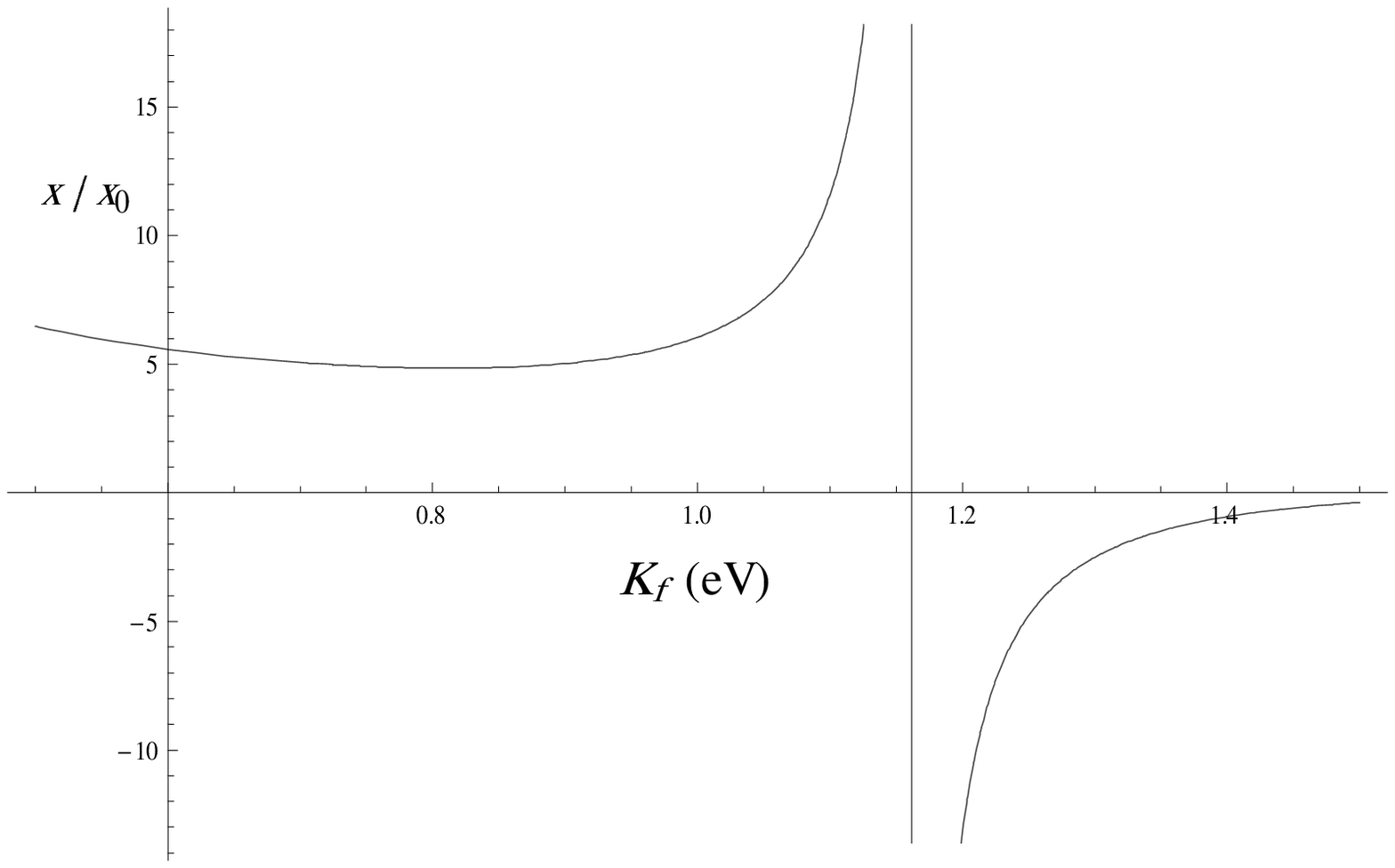} \caption{The magnetic
susceptibility of system at zero temperature as a function of fermi
momentum of system. As $k_f \longrightarrow \frac{m \alpha}{\pi}$
the $\chi$ tends to infinity and diverges. This means that the phase
transition is second order phase transition.} \label{f4}
\end{figure}

At zero temperature we can solve the Eq.(\ref{23}). The result is
shown in figure(\ref{f1}), there is a specific density (depended on
coupling constant), that for densities higher than it, the
ferromagnetism phase will appeared. This figure also shows that if
density becomes small the ferromagnetism phase can not exist inside
the system. At non-zero temperature we can solve the Eq.(\ref{22})
and find $m^*$ as a function of $p, T$ and density. Minimizing the
free energy with respected to $p$, at constant temperature, yields
to figures \ref{f2} and \ref{f3}.
 The fig [2] shows the density dependence of the polarization
parameter at various temperatures. Increasing the temperature leads
to disappearing the ferromagnetic phase. At low densities, also the
ferromagnetism disappeared. comparing the fermi energy of the system
with respected to the thermal energy of the system can help us to
find the reason. For states that their fermi energy is smaller than
thermal energy, the fermi distribution function reduced to the
Maxwell distribution and the difference between different spin
states ignorable. Figure \ref{f3} display the temperature dependence
of the polarization parameter at different densities. As we
expected, by increasing the temperature, the polarization of system
reduced and suddenly vanishes.
To obtain the magnetic susceptibility of the system we can use the
following relation \cite{tatsuminew}:
\begin{equation}
\chi^{-1}=\frac{1}{\rho^2 \mu_B}\frac{d^2 F}{d p^2}|_{p=0}
\label{29}
\end{equation}
At zero temperature the $p=0$ equivalent to $k_f = \frac{m
\alpha}{\pi}$. The magnetic susceptibility of the system is plotted
as a function of fermi momentum in figure \ref{f4}. It has a
divergence at $k_f= \frac{m \alpha}{\pi}$. This divergence is the
sign of second order phase transition inside the system.

\section{Summary}
The possibility of existence of the ferromagnetism phase inside the
electron gas investigated. The equations have written at low
temperature limit. However we can write the equations in general
form. we used the one photon exchange interaction at relativistic
region and then calculate the non-relativistic limit of interaction.
The spin dependence of the hamiltonian appears automatically.
According the results, the ferromagnetism phase can appeared at low
temperatures. The coupling constant has important rule in critical
density and temperature for phase transition. The magnetic
susceptibility at zero temperature calculated. The magnetic
susceptibility becomes infinite at $k_f\longrightarrow \frac{m
\pi}{\alpha}$, and this means that the phase transition is second
order. We also can obtain the critical exponents of the system by
expanding the magnetization and magnetic susceptibility of the
system near the critical point.
\section{Acknowledgment}
We would like to thanks Prof. M. V. Zverev for useful and important
guidance. We also would like to thank the Research Council of
University of Tehran and Institute for Research and Planning in
Higher Education for the grants provided for us.

\end{document}